\documentclass[aps,twocolumn,preprintnumbers,pra]{revtex4}
\usepackage{graphicx}
\usepackage{amsmath}
\usepackage{amssymb}
\usepackage{latexsym}
\usepackage{amsbsy}
\usepackage{array}
\usepackage{amsfonts}
\usepackage{mathrsfs}
\usepackage{color}
\usepackage{hyperref}
\usepackage{mathtools}
\usepackage{footnote}

\newcommand{\ket}[1]{|#1\rangle}

\newcommand{\be}{\begin{eqnarray}}
\newcommand{\ee}{\end{eqnarray}}

\begin{document}
\title{Detection efficiency loophole and  Pusey-Barrett-Rudolph theorem}
\author{Arijit Dutta, Marcin Paw\l owski, Marek \.Zukowski }
\affiliation{Institute of Theoretical Physics and Astrophysics, University of Gda\'{n}sk, 80-952 Gda\'{n}sk, Poland}

\received{\today}

\begin{abstract}
Detection efficiency loophole poses a significant problem for experimental tests of Bell inequalities. Recently discovered Pusey-Barrett-Rudolph (PBR) theorem suffers from the same vulnerability. In this paper we  calculate the critical detection efficiency, below which the  PBR argument for the ontic nature of quantum state is inconclusive. This is done for the maximally $\psi$-epistemic models. We use two different definitions of this property. The optimal number of parties, for which the critical detection efficiency is the lowest is given.  We also approach the problem from the opposite direction. We provide a function which enables us to specify which epistemic models are ruled out by the results of an experiment with a given detection efficiency.
\end{abstract}

\maketitle

\section{Introduction}

The status of the quantum state is a topic of a long standing discussion. There are many views on this issue in physics community.
The ensemble one: the quantum state is a theoretical description of a statistical ensemble of equivalently prepared systems and there are no underlying states for individual systems (see e.g. \cite{ORTHODOX}). Another school of thought, the {\it ontic} interpretation,  sees quantum state as a real physical object,  an inherent property of an individual quantum system. Underlying states exist, but they uniquely point to the quantum state. A competing  point of view, the {\it epistemic} one, considers,  the quantum state as a mere mathematical object to calculate probability of certain events. It is a state of knowledge, \cite{HR}. However, a more basic description of the system is possible involving real physical states of individual systems,  or hidden variables. The principal technical difference between the ontic and epistemic approach, is that in the later case two different,
 but non-orthogonal, quantum states may be linked with the same ``physical state".  Bell's Theorem tells us that a description with such variables cannot be local, however  it does not exclude the possibility of non-local theories.

The Pusey-Barrett-Rudolph (PBR) Theorem \cite{PBR}, is a major advance in the studies of the relation between ontic and epistemic views. It states that, one can find experimental situations, for which quantum mechanical predictions force us, if we allow hidden variables or states, to adopt the ontic interpretation of the quantum state. This result got much attention  and experiments followed  \cite{Miller,Patra,Patra1}.  However, like all  experiments probing  foundations of quantum mechanics the PBR proposal suffers from the detection efficiency loophole \cite{Fine}, i.e. the possibility that in the actual experiment the seemingly quantum effect is due to a post-selection of the events.  Only the  sub-sample of the cases in which all  detectors stations  register particles may be following the quantum predictions.  In fact the effect of  detection efficiency loophole  is much more severe in the case of PBR than for Bell inequalities. In the latter case, for any "relevant" inequality there exists a critical detection efficiency,  which,  if attained in  the experiment, rules out the local hidden variable description. In the former however the  efficiency must be  100\%. This makes PBR theorem not viable for experimental tests, see footnote \footnote{Of course theorems cannot be tested. They are logical statements. But
such tests have two-fold significance. They test whether the given
prediction of quantum mechanics, on which the theorem is based, agrees
with laboratory measurements,  and whether the phenomena required in the
theorem are experimentally observable (with current technology). If we
have unconditional positive answers in both cases, then the thesis of the
 theorem may be thought of as a law of nature, or an expression of
some law. Thus, in the end, we always test quantum mechanics, as a
theory of nature.},  unless some additional assumptions are made about hidden variables. Such assumptions lead to finite critical efficiencies. The purpose of this paper is to study the critical minimal  detection efficiency,  required for the PBR experiment to be conclusive,  under  certain reasonable conditions imposed  on the distributions of hidden variables.

First we present a short description of PBR argument. Next we discuss  detection efficiency loophole  in the context of Bell inequalities and explain why for PBR its effect is so  much stronger. Finally  we move to the main part of our paper. We find the critical detection efficiency for the following ``reasonable"  additional assumption: hidden variable  distributions associated with  two different quantum states are required to have the same overlap as the states. Even with this assumption we obtain a very high critical detection efficiency, below which experimental tests,  based on  PBR argument on ontic nature of quantum state,  are  inconclusive.

\section{PBR Theorem}

Let us give a short presentation of the results of  Pusey et al. \cite{PBR}.  Within the {\it non-orthodox view} allowing for hidden states or variables, the authors  have  shown a unique relationship  between hidden-physical reality and the quantum state.

Let's consider two different  preparation procedures  resulting in two quantum states $|\psi_1\rangle$ and $|\psi_2\rangle$. The states  are assumed to be different, but do not have to be orthogonal. The underlying  hidden-physical states  will be denoted by $\lambda$. One can assume  that there exists a probability distribution  $\rho_{\psi}(\lambda)$ in some  ontic  space $R$  associated with the given state $|\psi\rangle.$  If  supports of   $\rho_{\psi_1}(\lambda)$ and $\rho_{\psi_2}(\lambda)$ overlap, then there is at least one hidden-physical state common to both distributions. However, if the supports do not overlap, then they do not share any  common hidden-physical state. Pusey et al. in their argument have put forward  a gedanken-experiment, for which,  if underlying hidden variable model can explain the probabilistic nature of quantum mechanics, $\rho_{\psi_1}(\lambda)$ and $\rho_{\psi_2}(\lambda)$ must  have disjoint supports. This implies to the ontic nature of the state $|\psi\rangle,$ as hidden variables pinpoint the state with which they are associated.

\begin{figure}[h]
  \centering
  \includegraphics[width=0.9\linewidth]{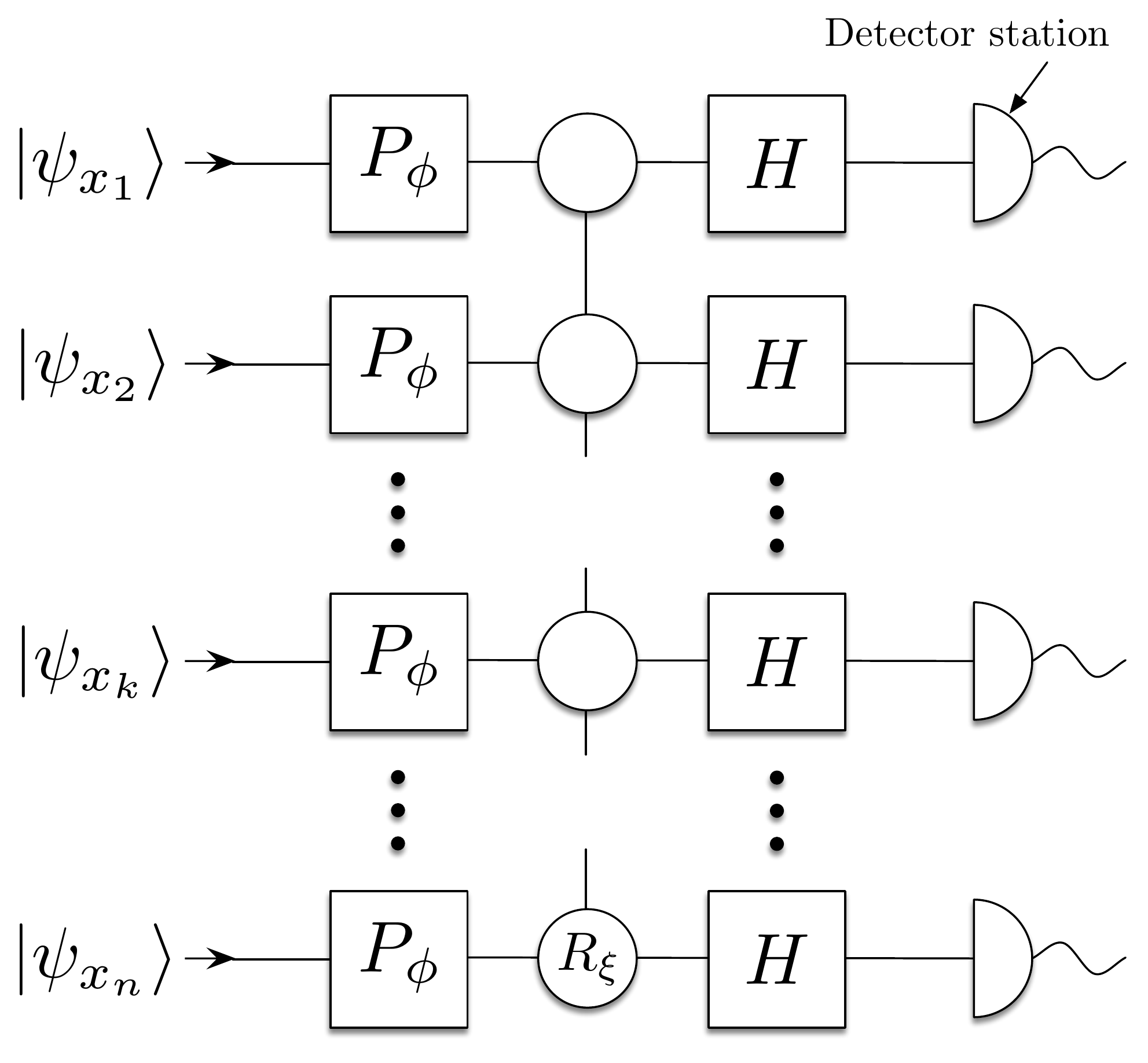}
\caption{PBR-argument is based on a joint measurement in a quantum circuit of n-qubits followed by a measurement on each qubit  in the computational basis. The single qubit gates are defined as $P_{\phi}=|0\rangle\langle0|+e^{i\phi}|1\rangle\langle 1|$  and the Hadamard gate,  $ |+\rangle\langle0|+|-\rangle\langle 1|$. The entangling gate in the middle rotates only  $|00\cdots 00\rangle$,  $R_{\xi}|00\cdots 00\rangle=e^{i\xi}|00\cdots 00\rangle.$ Each of the outputs is observed by a single detector station.}
\label{fig:PBRD}
\end{figure}

 In general, if measurement outcome depends on hidden-physical state  $\lambda,$ then for a given state $|\psi_1\rangle$  the   probability  to  obtain an outcome $|\psi_2\rangle$  for a measurement $M=|\psi_2\rangle\langle\psi_2|,$  associated with a response function $\xi_M(\psi_2|\lambda) $  reads
\begin{eqnarray}
\label{overlap111}
P(\psi_2|\psi_1)=\int_{R}\xi_M(\psi_2|\lambda)\rho_{{\psi}_1}(\lambda) d\lambda,
\end{eqnarray}
where $0\leq\xi_M(\psi_2|\lambda)\leq 1.$ If hidden variables can reproduce quantum mechanical predictions, then  by Born rule, $P(\psi_2|\psi_1)=|\langle\psi_2|\psi_1\rangle|^2.$
The authors have considered a specific joint measurement on a composite system with $n$ independently prepared sub-systems. They show that,  if probability distributions of hidden-physical states corresponding to two different quantum states overlap, the common  hidden-physical states from the  overlap region must  lead to measurement outcomes which are forbidden by quantum mechanical predictions. A diagram of a quantum circuit used in the gedanken-experiment \cite{PBR} is presented in  Fig.~$\ref{fig:PBRD}$.

The construction of ~\cite{PBR} runs as follows. Each subsystem is prepared under conditions which exclude any interdependence between the sub-systems (e.g., the are prepared in mutually remote locations, at the same moment of time, etc.).
The following $2^n $ possible states are considered :
\begin{equation}
\psi_{x_1, x_2,\cdots x_n}=|\psi_{x_1}\rangle\otimes|\psi_{x_2}\rangle\otimes...\otimes|\psi_{x_n}\rangle, \label{STATES}
 \end{equation}
 where $x_i\in{1, 2}$ and $|\psi_1\rangle=\cos\frac{\theta}{2}\ket{0}+\sin\frac{\theta}{2}\ket{1}$ and $|\psi_2\rangle=\cos\frac{\theta}{2}\ket{0}-\sin\frac{\theta}{2}\ket{1}.$

If there exists a hidden-physical state $\lambda$, for which $\rho_{\psi_1}(\lambda)>0$ and $\rho_{\psi_2}(\lambda)>0$ then there is a non-zero probability that $\psi_{x_1, x_2,\cdots x_n}$ will be prepared in the state $\lambda^{\otimes n}$ (this notation means: all systems in the hidden state $\lambda$) regardless of the choices of $x_i$'s. The measurement is chosen in such a way that, according to quantum predictions, for every choice of $x_1, x_2,\cdots x_n,$ one outcome, different for each $\psi_{x_1, x_2,\cdots x_n},$  is prohibited (i.e. its probability is zero). The number of possible outcomes is the same as the number of possible preparations. Thus, if the hidden variable  theory is to be in agreement with quantum mechanics then $\lambda^{\otimes n}$ must lead to probability zero for any outcome! Therefore, $\lambda^{\otimes n}$ can not be  the underlying state linked to any $\psi_{x_1, x_2,\cdots x_n}.$  We have  a contradiction. This implies that the premise, i.e existence of $\lambda$, for which $\rho_{\psi_1}(\lambda)>0$ and $\rho_{\psi_2}(\lambda)>0$ is wrong, which leads us to the conclusion that there is unique correspondence between any  hidden state $\lambda$ and the quantum state with which it is associated. That is the quantum state is of an {\em ontic}  nature.

However, this argument explicitly assumes that the particles are always detected. If there exists a mechanism,  which makes at least one of the detectors not click when the measurement is done on the hidden state $\lambda^{\otimes n},$ then there is no contradiction. This  is  the detection efficiency loophole. Since PBR argument does not provide us with any method  of estimating the probability with which $\lambda^{\otimes n}$ is generated,  we have no way of estimating how rarely  the detectors should fail to click to avoid this loophole. Therefore, extra assumptions are necessary.

\section{Bell's inequalities and detection loophole}

To experimentally refute the possibility of local, realistic description of quantum systems one has to violate a Bell inequality making sure that the  conditions used to derive them  are satisfied. Such conditions include the specific properties of the considered Bell experiment. If the experiment has specific features which are different than the ones assumed in derivation of Bell's inequalities, which lead to inconclusiveness of the experiment, then we talk about loopholes. One of them is the detection efficiency loophole \cite{Pearle}.

In \cite{Gisin} an explicit local model is given which  mimics quantum correlations for a singlet state and projective measurements, provided the detection efficiency is below $67\%.$ For every Bell inequality there exists a threshold minimal detection efficiency,  which is required,  if one wants to reject local-realism  by violating the inequality.  For the simplest case of two binary measurements on two-qubit entangled state the best known result is due to Eberhard \cite{eberhard} who showed that the critical detection efficiency for CH inequality\cite{CH} is $\frac{2}{3}$. For systems of higher dimension the threshold decreases exponentially with the dimension \cite{serge}. For multipartite Bell inequalities  Cabello et al. \cite{cabello1} have shown that a detection efficiency of $\frac{n}{2n-1}$ is both necessary and sufficient to violate $n$-partite Mermin \cite{Mermine} inequalities.

In the next section we discuss similar problems for the PBR gedanken-experiment.

\section{Detection efficiency loophole in PBR theorem}

Our aim is to find the critical detection efficiency below which PBR's argument on ontic nature of quantum states does not work anymore.

If ontic description is to be true, then each hidden-physical state must be uniquely linked to a {\it single} specific quantum state. If two different probability distributions of hidden-physical states  corresponding to two different quantum  states partially overlap, then there is an ambiguity in one to one relationship  between hidden-physical state and quantum state.  More precisely, there is a non zero probability that two different preparation methods corresponding to two quantum states may lead to same hidden-physical state.
If there are $n$ subsystems and one considers the $2^n$ possible  combinations of states (\ref{STATES}), for the epistemic approach there are  an underlying hidden-physical states  of the joint system $\lambda_0^{\oplus n}$, which can correspond to any of these. To save $\psi$-epistemic models from contradicting quantum predictions it suffices that one of the detectors used in the experiment fails  to  click whenever the measured compound system is in  $\lambda_0^{\oplus n}.$

Before we start our analysis we state that, we  assume, that  the detection inefficiency is the only experimental imperfection that we take into account.

Let $p$ be the probability that the supports of two probability distributions  $\rho_{\psi_1}(\lambda)$ and $\rho_{\psi_2}(\lambda)$ overlap. In the next section, we  estimate  $p$ as a measure of {\em epistemic overlap} (see  Ref~\cite{oje})  between two quantum states $|\psi_1\rangle$  and $|\psi_2\rangle.$  For simplicity we assume that the value of $p$  does not depend on the $n$-system quantum state  $ \psi_{x_1, x_2,\cdots x_n},$ which is prepared. Thus the total probability, associated with the overlap of two distributions  $\rho_{\psi_1}(\lambda),$  and $\rho_{\psi_2}(\lambda)$ for $n$ independently prepared subsystems in state $\psi_{x_1, x_2,\cdots x_n},$  is  $p_1=p^n$.

In the circuit shown in Fig.~$\ref{fig:PBRD}$ there are as many detector stations  as subsystems. Assume that the detectors  have detection efficiency $\eta$ and  their detection probabilities are independent. Then,  the probability that at least one of the detection stations registers no  click is $1-\eta^n$. If the detection loophole is to be solely responsible for the fact that no outcomes which contradict quantum mechanics are registered, then $p_1$ has to be lower or equal $1-\eta^n$. Thus, the critical detection efficiency is given by
\begin{eqnarray}
\label{efficiency}
&\eta=(1-p^n)^\frac{1}{n}.&
\end{eqnarray}

One can find in  Ref.\cite{PBR} that values of the angle $\theta,$ which defines $|\psi_1\rangle$ and $|\psi_2\rangle$, and their scalar product $\langle\psi_1|\psi_2\rangle=\cos\theta,$ determine how many qubits one has to have, so that the PBR argument for gedanken-experiment of Fig.~$\ref{fig:PBRD}$ can work. If one denotes the the number of qubits by $n,$ then the relation is given by

\begin{equation}
\label{overlap}
0<2 \arctan(2^{\frac{1}{n}}-1)\leq\theta< \frac{\pi}{2}.
\end{equation}
Thus we have a functional relation between the minimal value of $\theta$ in the gedanken-experiment, and the number of qubits:

\begin{equation}
\label{io}
n(\theta_\text{{min}})=\left\lceil \frac{1}{\log_2(1+\tan(\frac{\theta_\text{{min}}}{2}))}\right\rceil,
\end{equation}
Therefore, if one, under some assumptions, can fix $p$ (see e. g. next section) by combining $\eqref{efficiency}$ and $\eqref{io}$ one can get the minimal efficiency $\eta$ as a function of $\theta.$

Because PBR theorem does not say anything about how big $p$ can be,  it can be in principle arbitrarily small. Thus,  the critical efficiency given by (\ref{efficiency}) can  be arbitrarily close to 1. However this will not be the case if one assumes additionally some specific relation between $p$ and $\langle\psi_2|\psi_1\rangle.$

\section{Maximally $\psi$-epistemic models}

\begin{figure}[h]
  \centering
  \includegraphics[width=.9\linewidth]{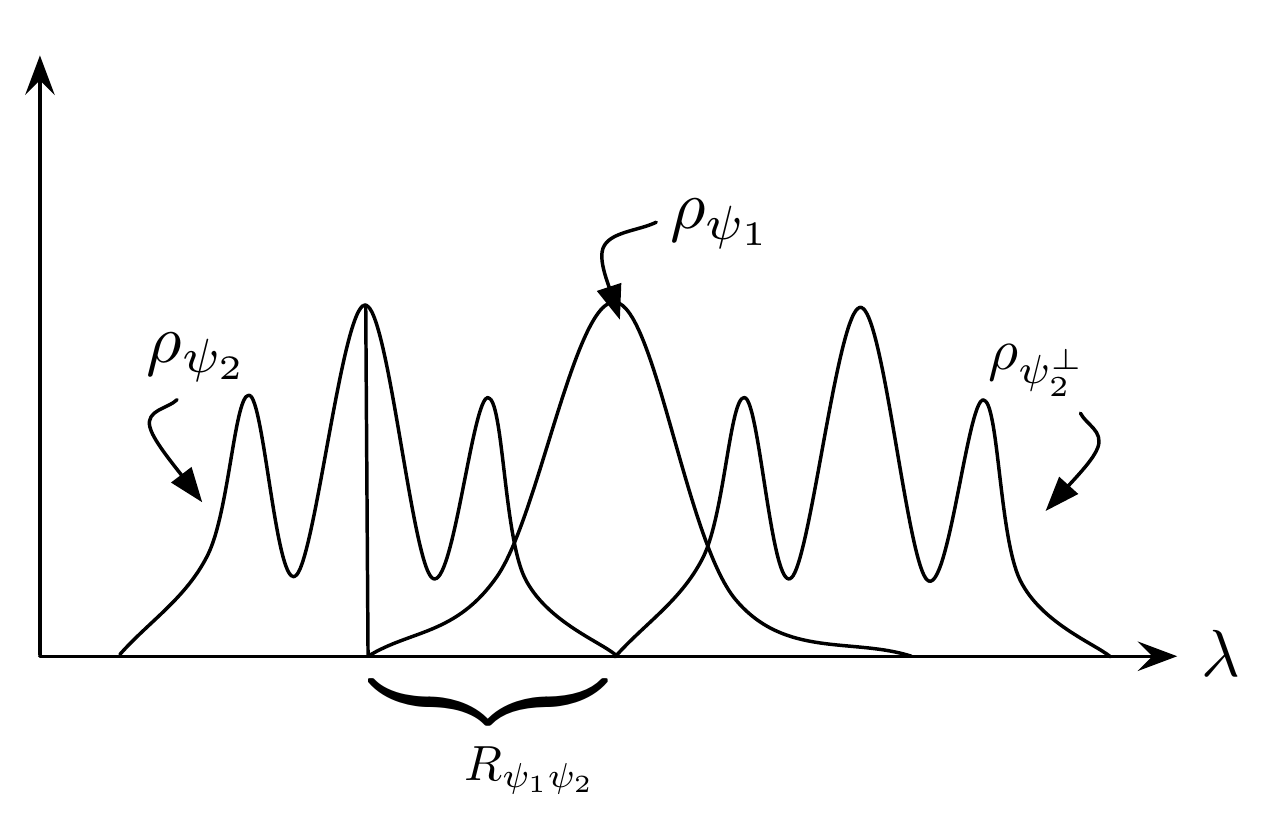}
	\caption{A schematic depiction of a maximally $\psi$-epistemic model. Different probability distributions are plotted on hidden variable space $\lambda$. Supports of $\rho_{{\psi}_2}$ and  $\rho_{{\psi}^\perp_2}$  do not overlap due to orthogonality of $\ket{\psi_2}$ and $\ket{\psi^\perp_2}$. $R_{\psi_1\psi_2}$ is the overlap region of $\rho_{\psi_2}$ and $\rho_{\psi_1}.$  In this class of models every $\lambda$ from the support of $\rho_{\psi_1}$ belongs either to the support of $\rho_{\psi_2}$ or $\rho_{\psi^\perp_2}$.  }
	\label{fig:ovverlap}
\end{figure}

The additional assumption that we now make is that the model that we are trying to falsify is maximally $\psi$ -epistemic \cite{Leifer}. By definition of Ref.~\cite{Leifer},  it implies that, see  Fig.~\ref{fig:ovverlap}

\begin{eqnarray}
\label{fig:overlap11}
 &p=\int_{R_{\psi_1\psi_2}}\rho_{{\psi}_1}(\lambda) d\lambda&\nonumber\\&=\int_{R_{\psi_1\psi_2}}\rho_{{\psi}_2}(\lambda) d\lambda=|\langle \psi_2|\psi_1 \rangle|^2=\cos^2\theta.&
\end{eqnarray}

For the maximally $\psi$-epistemic models  the union of the supports of any quantum  basis states  spans the whole space of hidden-physical states. Note that this implies that in the case of a two dimensional subspace, any pair of orthogonal states has the same union of supports. Thus we  necessarily have non-zero overlaps for $\rho_{\psi_1}$ and $\rho_{\psi_2},$  see  Fig.~\ref{fig:ovverlap}, provided are linked with two non-orthogonal states.

\subsection{Alternative approaches}

The validity of the PBR argument rests crucially on {\it preparation independence} assumption, which means that the probability distributions of $\lambda$'s corresponding to two different systems are independent. Recently there have been several attempts to re-derive PBR result without this assumption but with limited success \cite{oje,n1,n2,n3}. While PBR argument works for any pair of states, results of \cite{oje,n1,n2,n3} are not so general. The states for which these arguments hold are of a certain dimension (at least 3 in all the cases) and it is known that without taking additional assumptions it is impossible to rule out an ontological model for qubits \cite{terry}. The results of \cite{n2} apply only to Hadamard states and the results from \cite{n1,n3} to PP-incompatible \cite{PPi} ones.

In all these works only models close to maximally $\psi$-epistemic are refuted. The degree of ``closeness"  is treated differently in the  papers. E. g. in \cite{oje} it is based on (\ref{fig:overlap11}) and parameterized by  $\Omega$, which is between $0$ and $1.$  The relation reads
\be \label{pom}
p=\int_{R_{\psi_1\psi_2}}\rho_{{\psi}_1}(\lambda) d\lambda=\Omega |\langle \psi_2|\psi_1 \rangle|^2=\Omega \cos^2\theta.
\ee
The  maximally $\psi$-epistemic case  corresponds to $\Omega=1$ \cite{Leifer}. We call these models $\psi_{\Omega}$-epistemic.

In \cite{n1,n2,n3} the overlap of probability distributions is measured differently. One can introduce another parameter $k,$ again $0<k<1$  such that
\be
\int \min\{\rho_{{\psi}_1}(\lambda),\rho_{{\psi}_2}(\lambda)\}d\lambda=k(1-\sin\theta).
\ee
It's relation with $p$ is given  by
\be \label{pk}
p>k(1-\sin\theta).
\ee
One can now get another definition of a maximally $\psi$-epistemic model by considering one with $k=1$. We call such models $\psi_{k}$-epistemic.

\section{Critical detection efficiency for maximally $\psi$-epistemic models}

\begin{figure}[h]
  \centering
  \includegraphics[width=.8\linewidth]{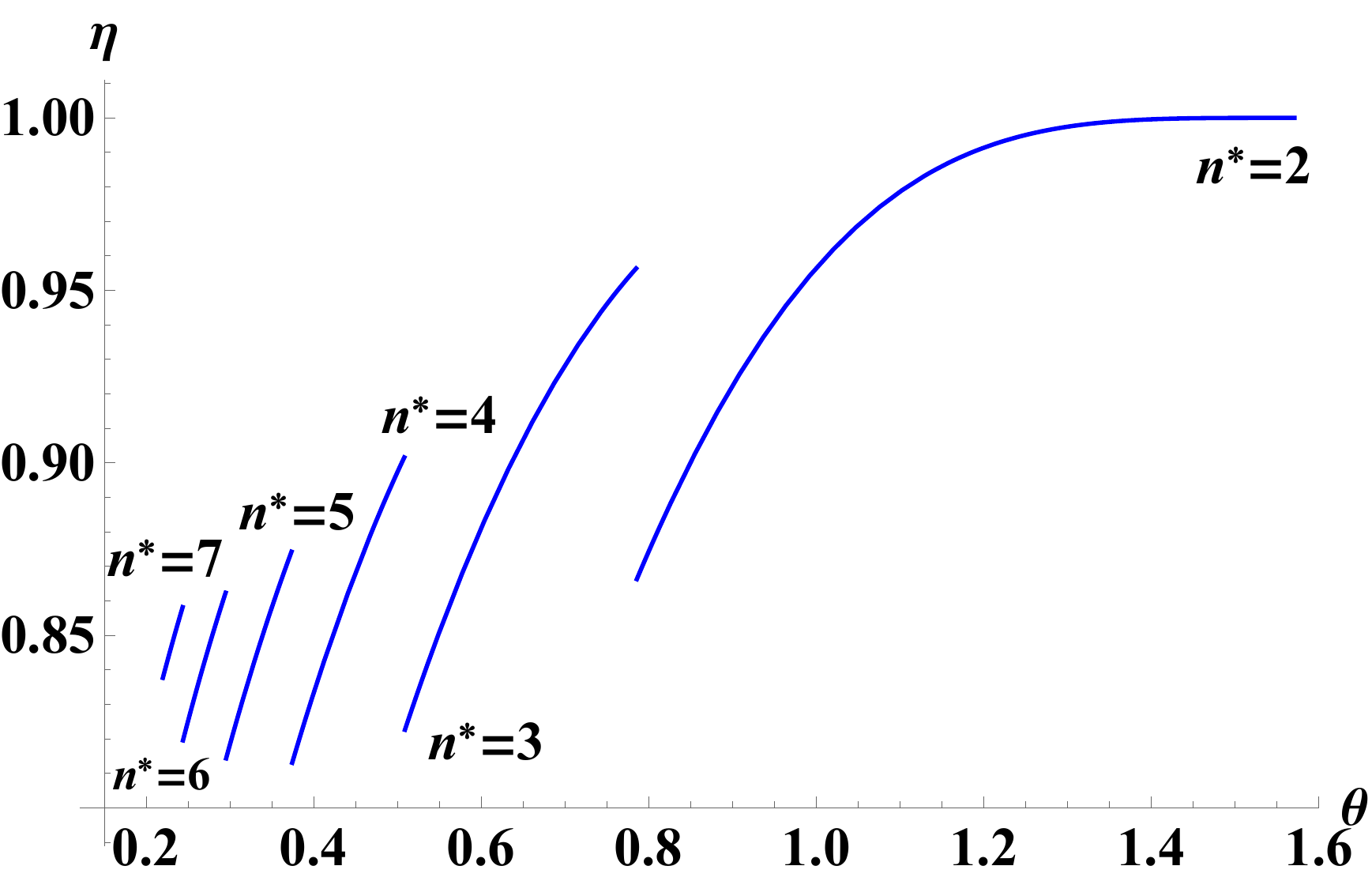}
	\caption{Dependence of critical detection efficiency $\eta$ as a function of $\theta$  measured in radians for different values of $n^* \in \{2, 3,\cdots 7\}$ for tests of maximally $\psi_{\Omega}$-epistemic models. }
	\label{fig:n2c}
\end{figure}

\begin{figure}[h]
  \centering
  \includegraphics[width=.8\linewidth]{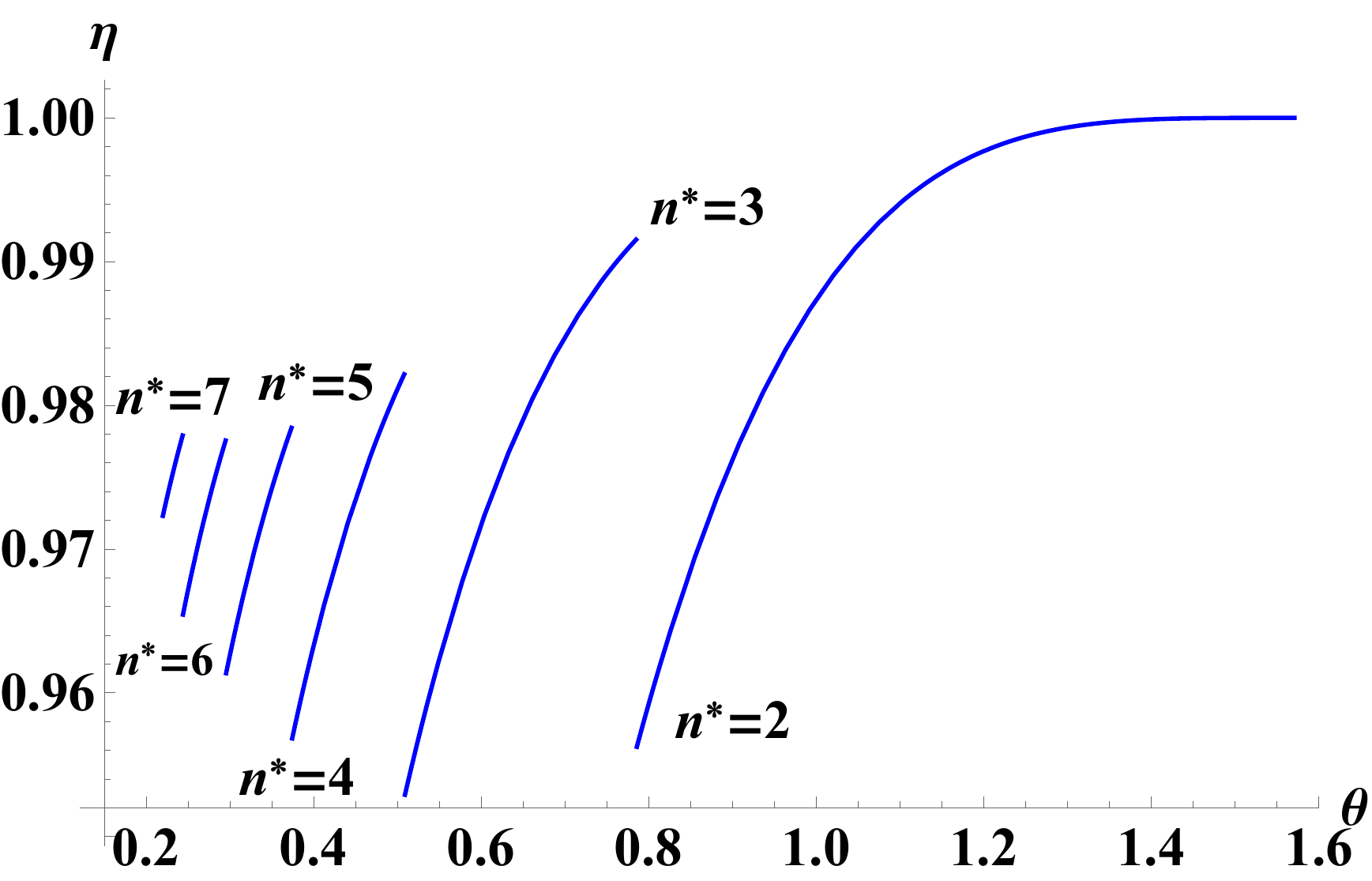}
	\caption{Dependence of critical detection efficiency $\eta$ as a function of $\theta$  measured in radians for different values of $n^* \in \{2, 3,\cdots 7\}$ for tests of maximally $\psi_{k}$-epistemic models.}
	\label{fig:n2d}
\end{figure}

By plugging either (\ref{pom}) or (\ref{pk}) and (\ref{io}) into (\ref{efficiency}) and taking $\Omega=1$ or $k=1$, respectively, we obtain the critical value of detection efficiency as a function of $\theta$ only. The results are  plotted in Fig. $\ref{fig:n2c}$ for $\psi_{\Omega}$-epistemic and in Fig. $\ref{fig:n2d}$ for $\psi_{k}$-epistemic models. The discontinuities in the graphs are due changes of the number of particle which is optimal for the given range of $\theta$, see formula  (\ref{io}). It is easy to notice that the choice of $\theta$ which leads to the lowest detection efficiencies close to the point at which one has to change the number of qubits involved  i.e. when $n^*=\frac{1}{\log_2(1+tan(\frac{\theta_{\text{min}}}{2}))}$ is an integer.

We can also find the critical detection efficiency as a function of number of subsystems by taking $\theta$'s which correspond to integer values of $n^*$. These values are given in Table~\ref{table}. The minimal critical detection efficiency for testing maximally $\psi_{\Omega}$-epistemic models is 81.3\% and it is reached for an experiment with {\em four} subsystems while for testing maximally $\psi_{k}$-epistemic ones it is 95.3\% and it requires {\em three} subsystems.

\begin{table}[t]
\centering
\begin{tabular}{cccc}
\hline\hline
$\theta_{\text{min}}$(in radians) & $n^*$ & $\eta_{\Omega}$ & $\eta_{k}$\\
\hline
 0.785 & 2 & 0.866 & 0.956\\
 0.509 & 3 & 0.822 & 0.953 \\
 0.374 & 4 & 0.813 & 0.957\\
 0.295 & 5 & 0.814  & 0.961\\
 0.244 & 6 & 0.819 & 0.965\\
0.207 & 7 & 0.826 & 0.969\\
\hline\hline
\end{tabular}
\caption{Critical detection efficiencies for different number of subsystems ($n^*$). $\eta_{\Omega}$ and $\eta_{k}$ correspond to maximally $\psi_{\Omega}$-epistemic and $\psi_{k}$-epistemic models respectively.}
\label{table}
\end{table}

\section{Non-maximally $\psi$-epistemic models}

\begin{figure}[h]
  \centering
  \includegraphics[width=.7\linewidth]{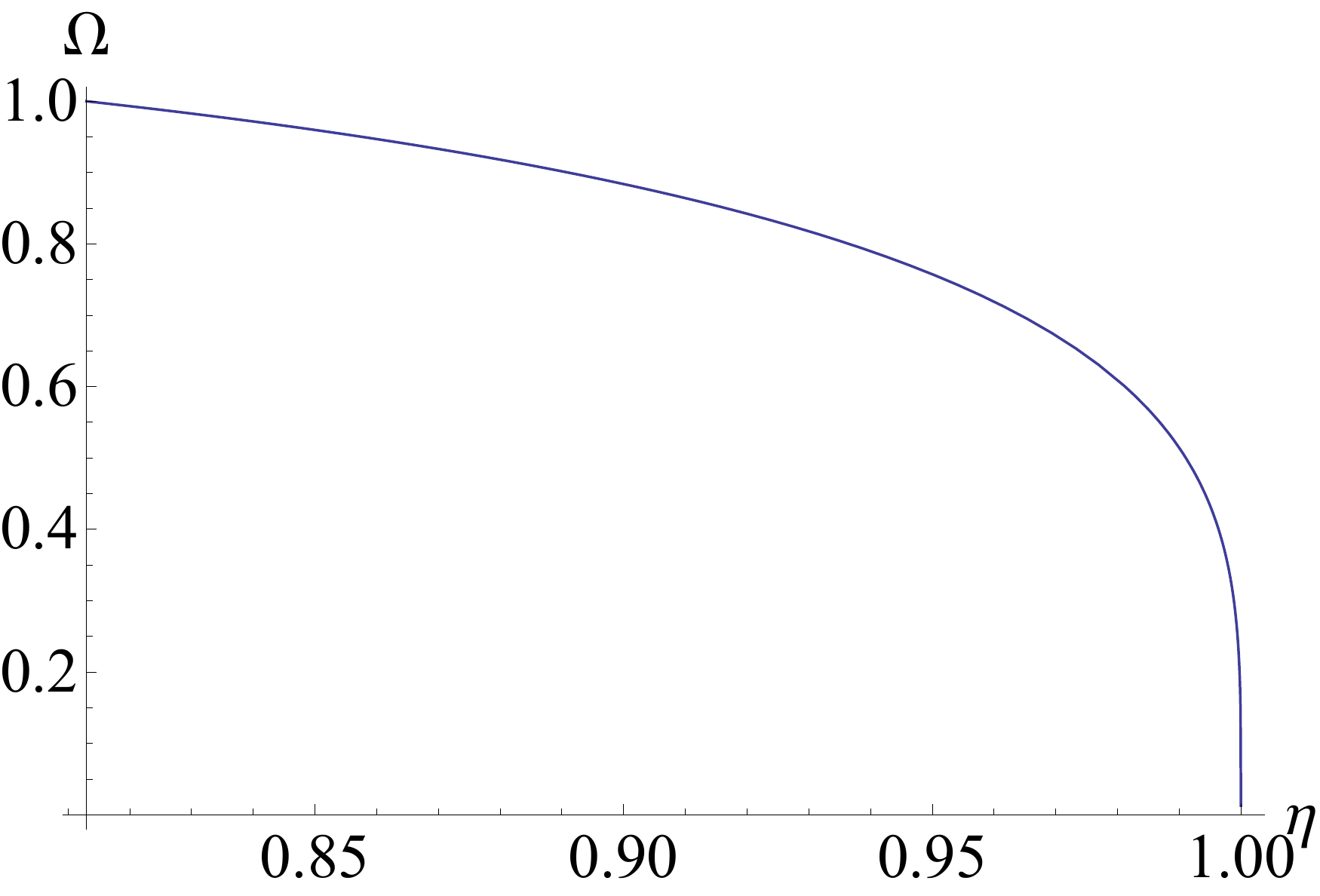}
	\caption{Critical value of $\Omega$ as a function of detection efficiency $\eta$. For this plot optimal values of the parameters $n^*$ and $\theta_{\text{min}}$ have been used, ie. $n^*=4$ and $\theta_{\text{min}}=0.374$.}
	\label{fig:pom}
\end{figure}

\begin{figure}[h]
  \centering
  \includegraphics[width=.7\linewidth]{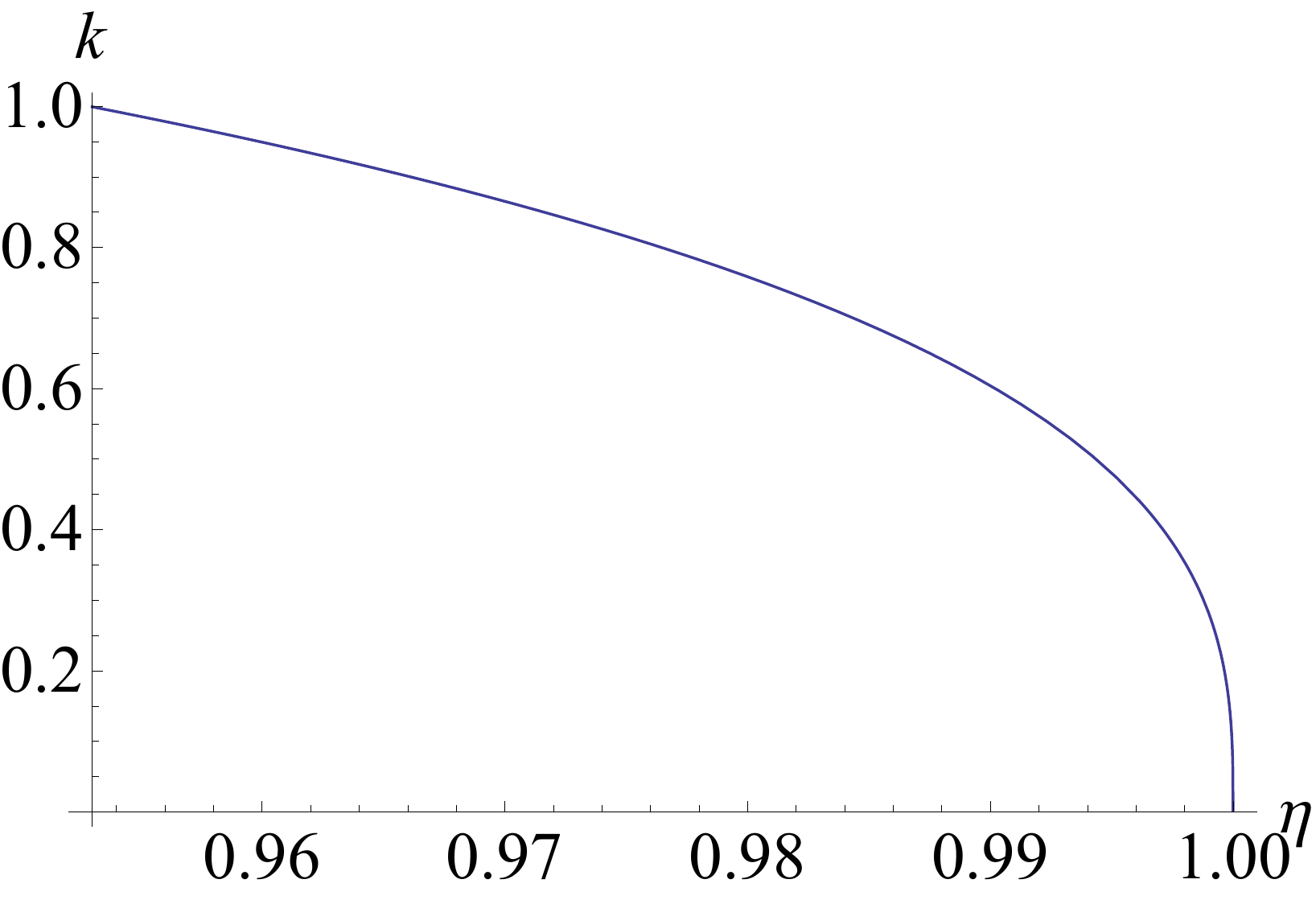}
	\caption{Critical value of $k$ as a function of detection efficiency $\eta$. For this plot optimal values of the parameters $n^*$ and $\theta_{\text{min}}$ have been used, ie. $n^*=3$ and $\theta_{\text{min}}=0.509$.}
	\label{fig:pk}
\end{figure}

Our analysis can be applied also for non-maximally $\psi$-epistemic models. It suffices to  plug (\ref{pom}) or (\ref{pk}) and (\ref{io}) into (\ref{efficiency})  and take any value of $\Omega$ or $k$ between 0 and 1. This may be used for checking for which values of $\Omega$ and $k$ there can be reasonable detection efficiency for a experimental test. Figs. \ref{fig:pom} and \ref{fig:pk} show the dependence of critical $\Omega$ and $k$ as a function of the detection efficiency. Notice that around values of $\Omega\approx 0.7$ and $k\approx 0.9$, the required efficiency starts to to be prohibitively high, as it reaches $97\%$, approximately the current state of the art.

\section{Conclusions}

We study the detection efficiency loophole in the context of PBR theorem. We point out that without additional assumptions the theorem only holds in the ideal case. For the non-deal case of inefficient detectors for  maximally $\psi$-epistemic models, we  obtain critical detection efficiency of 81.3\%. If one uses a different definition of epistemicity, of the $\psi_k$ type, the threshold increases to 95.3\%. It is worth noting that this value is reached neither by the simplest case of {\em two}  subsystems or in the limit of infinitely many but by an intermediate number ({\em three} in the fist case and {\em four} in the second).

Our results show that the detection efficiency thresholds for quantum test based on the  PBR gedanken-experiments are much higher than in the case of Bell inequalities. Not only it makes inconclusive possible experimental tests  (see footnote~[22]) which are done and analyzed without any other additional assumptions, but also it shows that  if the strongest additional assumption concerning epistemicity is made the obtained critical value of detection efficiency is very high. Yet, it is almost within reach of the current state-of-the-art technology and we hope a loophole free refutation of maximally $\psi$-epistemic models would be performed soon.

\acknowledgements

 AD would like to  thank Matt Leifer  for stimulating discussions and Junghee Ryu for helping with figures. AD is supported by the International PhD Project "Physics of future quantum-based information technologies": grant: MPD/2009-3/4 of Foundation for Polish Science.  M.P. is supported by FNP program TEAM, ERC grant QOLAPS and NCN grant 2012/05/E/ST2/02352.

\end{document}